\newcommand{\beq}{\begin{equation}}
\newcommand{\eeq}{\end{equation}}
\newcommand{\beqa}{\begin{eqnarray}}
\newcommand{\eeqa}{\end{eqnarray}}

\renewcommand{\lambda}{\ell}

\documentstyle[aps,prl,multicol,fancyheadings,epsfig,epsf,amsbsy,amssymb,amstex]{revtex}
\begin{document}
\title{Inhomogeneity-Induced Superconductivity? }
\author{J. Eroles$^{1,2}$, G. Ortiz$^{1}$, A. V. Balatsky$^{1}$ and A.
R. Bishop$^{1}$}
\address{$^1$Theoretical Division, 
Los Alamos National Laboratory, Los Alamos, NM 87545, USA.\\
$^2$Centro At\'{o}mico Bariloche and Instituto Balseiro, 
S. C. de Bariloche, Argentina.\\}
\date{Received \today }

\maketitle

\begin{abstract}
A $t$-$J$-like model for inhomogeneous superconductivity of cuprate
oxides is presented, in which local anisotropic magnetic terms are
essential. We show that this model predicts pairing, consistent with
experiments, and argue how the macroscopic phase-coherent state
gradually grows upon lowering of the temperature. We show that
appropriate inhomogeneities are essential in order to have significant
pair binding in the thermodynamic limit. Particularly, {\it local}
breaking of $SU(2)$ spin symmetry is an efficient mechanism for
inducing pairing of two holes, as well as explaining the magnetic
scattering properties. We also discuss the connection of the resulting
inhomogeneity-induced superconductivity to recent experimental evidence
for a linear relation between  magnetic incommensurability and the
superconducting transition temperature, as a function of doping.

\end{abstract}
\pacs{Pacs Numbers: 74.20.-z, 74.20.Mn, 74.72.-h, 71.27.+a}

\vspace*{-1.0cm}
\begin{multicols}{2}

\columnseprule 0pt


There is a growing body of experimental evidence suggesting that the
superconducting state in cuprate oxides is ``inhomogeneous,'' such that the
locally defined charge density varies across the sample in the ground
state. Spatially inhomogeneous features in the spin and charge channels
have been indicated in a number of experiments on high-$T_c$ materials
\cite{exp1,Mook,Yamada,BB,noda,uchida}.
The simplest realization of this state is the so-called ``stripe''
phase where charges cluster in nanoscale linear patterns and the
remainder of the sample is essentially an antiferromagnetically
correlated insulator. This represents a nanoscale distribution of
charge and spin, rather than a global phase separation. These
experiments lead us to a central question: Is the
superconducting state found in high-$T_c$ cuprates inhomogeneous as a
result of spin/charge inhomogeneities? We believe that the answer to
this question is yes. Moreover, we argue that spatial spin/charge
inhomogeneities are in fact {\it necessary} for pairing and subsequent
formation of the superconducting state in these compounds. This
situation should be contrasted with the case of conventional
superconductors, resolved by the BCS (Bardeen-Cooper-Schrieffer)
theory, that starts with a homogeneous metallic state and describes the
formation of a homogeneous superconducting state. It is commonly
believed that magnetic correlations, characterized by the spin exchange
energy $J \sim 1500 K$ are responsible for the pairing interactions in
the cuprates and are, therefore, crucial for our inhomogeneous exchange
approach. Moreover, the existence of a spin gap has been 
experimentally proven \cite{spingap}. A model that naturally 
incorporates these features (inhomogeneities, magnetism and spin gap) 
is a $t$-$J$-like model with explicitly broken spatial and magnetic 
symmetries.

Of central importance for the present work is a microscopic model which
captures the main low energy physics of doped antiferromagnetic (AF)
Mott insulators. In particular, we show that our minimal model properly
describes the magnetic properties observed in a wide variety of doped
cuprate oxide materials. The key ingredient is the existence of
magnetic perturbations which explicitly break {\it local}
spin-rotational invariance (e.g., due to local spin-orbit coupling
\cite{nick}) and thereby induce substantial hole pair binding. We then
develop a mean-field theory of superconductivity based upon a
phenomenology from our microscopic model. We emphasize that in our
approach there are two, in principle different, energy scales; one
associated to the pairing of holes and another related to the phase
coherence of the pairs (that establishes $T_c$). Basically the
inhomogeneities induce a strong hole pairing, which in turn
Josephson-tunnel coherently between stripes, separated by insulating AF
regions, phase-locking into a macroscopic supercurrent superfluid
stiffness.  Recently, a simple linear relation between the
superconducting transition temperature $T_c$ and the AF
incommensuration $\delta$ has been observed for the LSCO \cite{Yamada}
and YBCO \cite{BB} high-$T_c$ compounds: $k_B T_c \propto \delta$,
where $k_B$ is the Boltzman constant. We find that for this relation to
hold we need a power law Josephson  tunneling. 

{\it Microscopic Model}.  Our model Hamiltonian describing the low
energy dynamics of CuO$_2$ planes is $H = H_{t\!-\!J} + H_{\rm inh}$,
where the background Hamiltonian $H_{t\!-\!J}$ is the standard $t$-$J$
model  

\begin{equation} 
\label{H} 
H_{t\!-\!J} = -t  \sum_{\langle i,j \rangle, \sigma} c^{\dagger}_{i
\sigma} c^{\;}_{j \sigma} + J \sum_{\langle i,j \rangle} ({\bf S}_i
\cdot {\bf S}_j - \frac{1}{4} \bar{n}_i \bar{n}_j ) \ . 
\end{equation} 

\noindent For the inhomogeneous component, we take 

\begin{equation} 
H_{\rm inh} = \sum_{\langle \alpha,\beta \rangle} \delta J_z \
S^z_\alpha S^z_\beta + \frac{\delta J_{\perp}}{2} \left( S^+_\alpha 
S^-_\beta + S^-_\alpha S^+_\beta \right)
\nonumber \ , 
\end{equation} 

\noindent with $\delta J_{\perp} \neq \delta J_z$, representing the
magnetic  perturbation of a static local Ising anisotropy, locally
lowering spin symmetry. Here $\langle i,j \rangle$ are near-neighbor
sites, while $\langle \alpha,\beta \rangle$ are two near-neighbor sites
characterizing the bonds that are perturbed and where $SU(2)$
spin-rotational invariance is explicitly broken. The network of
perturbed bonds form mesoscopic patterns determined by the distribution
of stripe segments. The spin-$\frac{1}{2}$ operator ${\bf S}_i =
\frac{1}{2} c^{\dagger}_{i \sigma} \mbox{\boldmath $\tau$}^{\;}_{\!\!
\sigma \sigma'} c^{\;}_{i \sigma'}$, the electron occupation number
$\bar{n}_i = c^{\dagger}_{i \uparrow} c^{\;}_{i \uparrow} +
c^{\dagger}_{i \downarrow} c^{\;}_{i \downarrow}$, and $c^{\dagger}_{i
\sigma}(c^{\;}_{i \sigma})$ creates(annihilates) an electron of spin
$\sigma$ in a Wannier orbital centered at site $i$; $\mbox{\boldmath
$\tau$}$ are the 3 Pauli matrices. This is a three-state model with the
hopping constrained to the subspace with no doubly occupied sites. In
the following, all energies will be measured in units of $J$. 


Our modeling strategy consists in assuming the existence of an 
inhomogeneous mesoscopic skeleton of stripe segments\cite{krun}, and then
exploring its consequences, mainly the competition between magnetism
and superconductivity. We do not address here the important problem of
the formation and stability of this skeleton morphology. The origin(s)
of ``stripe segment'' formation in high temperature superconductors is
as yet unclear and several physical mechanisms could act cooperatively
and be responsible for the generation of multiple length scales, among
them: spin-orbit coupling, local Jahn-Teller distortions induced by the
hole, effective interactions coming from a multi-band Hubbard Model (HM)
(including explicitly the oxygen and copper bands), oxygen buckling at
the stripe, and other local magnetoelastic effects \cite{yu}.
Competitions between attractive short range forces and repulsive long
range ones can certainly spontaneously break translational and/or
rotational invariance in the CuO$_2$ planes \cite{theory,theory1}, but
this is not necessarily the only mechanism. However, we show below that
the mere existence of appropriate {\it local} magnetic anisotropies is
crucial for pair-formation. 

We start by showing that, as far as we could numerically determine,
only by including a {\it local Ising} perturbation such as $H_{\rm
inh}$ in Eq. \ref{H} can a strong pairing of holes be obtained (see
Fig. \ref{fig_binding}). All the calculations were made using exact
diagonalization in small clusters with periodic boundary conditions in 
all spatial directions. We studied one-dimensional (1D) chains up to 16
sites and $8 \times 2$ clusters. Hereafter, we will view our clusters as
simulating systems where the longer direction is perpendicular to the
stripes. We investigated the system size scaling for the binding energy
of two holes defined as $E_b=(E_{2 \, {\rm holes}}-E_{0 \, {\rm
hole}})-2 (E_{1 \, {\rm hole}}-E_{0 \, {\rm hole}})$ for several models
in 1D and 2D. Although for small enough systems the binding energies
could be very large, they all seem to extrapolate towards no (or
extremely small) binding in the thermodynamic limit, with the clear
exception of the inhomogeneous $t$-$J$$J_{z}$ case. We have also
studied several one-band HMs, but we could again not find definite
binding. The $t$-$J$$J_{z}$ model, the only one unambiguously giving
binding in the thermodynamic limit, is obtained by breaking
spin-rotation symmetry in $d$ near-neighbor bonds $\langle \alpha,\beta
\rangle$, repeated with period $P$, by an amount $\delta J_z = 0$,
$\delta J_{\perp}<0$ in Eq. \ref{H}. This $t$-$J$$J_{z}$ model is a
most natural way to induce a spin-gap. We have checked that the
spin-gap is present for our $t$-$J$$J_{z}$ model\cite{fig5S}. The
inhomogeneities forming the superstructure, which we impose by hand in
the Hamiltonian, we term {\it stripes}. In Fig. \ref{fig_noni} we show
the hole correlation function $\langle g |n_0 . n_i| g \rangle$, where
$| g \rangle$ is the ground state of the system ($\langle g | g \rangle
= 1$). This correlation function gives information about the structure
of the pair. It can be seen that as the hopping strength $t$ is
increased beyond a characteristic value the second hole jumps from one
stripe to the neighboring one, starting from an initial configuration
where both holes are in the same stripe for small $t$. This can be
understood as a result of a length(time)-scale competition: the pair
size exceeds the stripe width. 

To explore the nature of the binding, we have examined the canonical  
transformation of a $t$-$J$ model from a one-band HM and traced  what
kind of perturbations would produce a $t$-$J$$J_{z}$ term. This 
corresponds to a term like $- V \sum_{\sigma} \bar{n}_{i,\sigma}
\bar{n}_{j,\bar{\sigma}}$ ($i,j$ first neighbors) in an extended HM,
which may in turn arise, for example, from local magnetoelastic (e.g.
oxygen (un)buckling) \cite{magneto} or spin-orbit couplings. Note,
again, that here this is a perturbation {\it only} at the stripes. This
kind of anisotropy manifests itself in two different ways in the
$t$-$J$$J_z$ model, enhancing both the $-\frac{1}{4} \bar{n}_i
\bar{n}_j$ and the easy axis (Ising) terms of the Hamiltonian. The
first one is an explicit pairing term for {\it electrons}. To see the
relative importance of each term we have calculated the binding energy
of a Hamiltonian like (\ref{H}) but excluding  the $-\frac{1}{4}
\bar{n}_i \bar{n}_j$ term and including bonds with broken
spin-rotational symmetry. This model corresponds to holes (with no
spin) propagating in an antiferromagnet, but not derived from a
canonical transformation of a one-band HM. We find that it still has
binding, as should be expected. Thus, the easy axis exchange term is
partially responsible for the binding energy.

In order to understand this exchanged-based pairing mechanism, it is 
useful to explore some limiting cases. When the magnetic energy scales 
are the most relevant ones: $(J_z = J + \delta J_z \ , \ J_{\perp} = J
+ \delta J_{\perp}) \gg t$, it is easy to realize that, depending upon
$J_z, J_{\perp}$ being smaller or larger than $J$, the holes will
prefer to be in the stripes or between stripes (with no binding),
respectively. The opposite limit, i.e. purely kinetic energy, leads to
delocalized holes and no binding. The situation where $J_z < J$ and $t$
is relevant corresponds to the intermediate regime where pairing is
observed. Notice that pairing of holes does not necessarily imply that 
holes should share the same stripe, they can occupy neighboring ones 
(see Fig. \ref{fig_noni}, upper panel), thus avoiding phase separation.
Details of the charge confinement and pairing potentials from the
(dynamic) spin-field profiles in the superlattice skeletons will be
given elsewhere. 


Having demonstrated a minimal model for hole binding, we have computed
spin correlation functions in clusters of size $N_x\times N_y = N$
($N_x=8, N_y=2$). Here, we simulate the stripes by including an
anisotropic $\delta J_{\perp}<0$ in one $y$-bond with $P=4$; the rest
of the bonds, including all the $x$-bonds, were not changed from the
background $t$-$J$ model (see inset, Fig. \ref{fig_sq}). We cannot
perform scaling on this size of inhomogeneous system, but the binding
energy is still considerable. We have included up to 6 holes. In the
case of four holes (the one more relevant to the stripes in the
underdoped regime for cuprate oxides) and small $t \ (\lesssim J)$ the
holes bind in pairs on each site of the inhomogeneous bond (see Fig.
\ref{fig_noni}, lower panel). 
 

In Fig. \ref{fig_sq} we show the spin-structure factor function
$S({\bold{k}}=(k_x,k_y))$ defined as: $\langle S_{\bold{k}}
S_{\bold{-k}} \rangle = (1/N) \sum_{\bold{i},\bold{j}} \exp(i\bold{k}.
\bold{r}_{\bold{i}}) \  \langle {\it g}| \bold{S_j}.\bold{S}_{\bold{j}
+ \bold{i}}| {\it g} \rangle$. This function corresponds to the
observable in the elastic neutron scattering experiments. For $t$
small, only one peak occurs with $k_y=\pi$, corresponding to two
essentially uncorrelated AF domains, isolated from each other by the
pinned hole wall. As $t$ increases ($t/J \gtrsim 2$, near the accepted
set of values of the 2D $t$-$J$ model for cuprates \cite{armando}), the
holes gain kinetic energy by visiting the first neighbor sites around
the anisotropy region, but still bind together. The effective width of
the pair thus increases to two sites. Magnetic energy is then gained if
the two domains shift their staggered magnetization by $\pi$. We
suggest that these ${\cal O}(t^2)$ processes are responsible for the
incommensuration ($\delta$) in $S(\bold{k})$ observed in the
experiments\cite{exp1}. This $\delta$ is the inverse of twice the
period $P$ of the stripes. In this picture the incommensuration is a
consequence of the holes and their kinetic energy, and is a property of
the ground state. Basically, it results from the competition between
hole delocalization and magnetic fluctuations. This contrasts with some
other proposed explanations, where $\delta$ is a magnetic thermodynamic
property \cite{birgeno}. It is interesting to note that this
incommensuration arises even in the homogeneous $t$-$J$ model although
for different values of $t$. This suggests that the experimentally
observed magnetic properties are already present in a homogeneous
$t$-$J$ model, but in order to obtain binding of holes appropiate
inhomogeneous terms must be included.

When more than four holes are added to the system, but only two bonds
are perturbed, $S({\bold{k}})$ changes qualitatively. Instead of
showing an incommensurability around $\bold{k}= {\bf Q} =(\pi,\pi)$, it
has a broad peak at $\bold{k}=(0,\pi)$. In this case the extra holes
are delocalized in the middle of the AF space between stripes. This
suggests that when the stripes reach their minimum separation, extra
holes are responsible for the experimental increase and ultimate
disappearance of the incommensurability.


{\it Model of Josephson Spaghetti.} It is important to relate the above
discussion to the experimental evidence for the incommensurate neutron
scattering peak, seen in LSCO (e.g., \cite{Yamada}) and YBCO compounds
\cite{Mook}. In both of these cases a simple {\em linear} relation
between $T_c$ and the peak incommensuration $\delta$ near ${\bf Q}$ (or
peak width in YBCO) is obeyed \cite{Yamada,BB}. Namely, 
\beqa 
k_B T_c = \hbar v^* \delta  \ . 
\label{v*} 
\eeqa 
The anomalously low velocity values for $v^*$ depend on the compound 
\cite{BB}. These velocities are independent of the carrier
concentration and the only doping ($x$) dependence entering Eq.
\ref{v*} is through $\delta(x)$. 

An interpretation of this relation is to connect possible
superconductivity mechanisms to the existence of the fluctuating
stripes. Here we focus on the simple proportionality between $T_c(x)$
and a doping dependent length $\ell(x)$, determined from the neutron
scattering: $T_c(x) \propto 1/\ell(x), \; \ell(x)  = 1/\delta(x)$. We
consider how the Josephson tunneling of pairs between stripe segments
can produce the relation between the phase ordering transition
temperature $T_c$ and the typical length $\ell(x)$. The stripe-stripe
distance $r$ is a random quantity due to intrinsic mechanisms as well
as disorder and/or crystal imperfections \cite{theory,theory1}.
Therefore, we will assume that the mean-field transition temperature
depends upon the Josephson coupling $\langle J(r) \rangle$, averaged
with some probability distribution of stripe separations. 

Our model Hamiltonian of random stripe separation and associated inter-
and intra-stripe random Josephson coupling (see Fig. \ref{spaghetti})
is  
\beqa
{\cal H} = \sum_{ij} J_{ij} \ \exp[i(\phi_i - \phi_j)] \ , 
\nonumber\\ 
J_{ij} = J(r_{ij}) = t_0/r_{ij}^{\alpha} \ , 
\label{ham} 
\eeqa 
where the summation is taken over the coarse-grained regions $i=
1,\cdots,{\cal N}$ with well-defined phases, labeled $\phi_i =
\phi(r_i)$ and $J_{ij}$ becomes zero eventually at large distances. 
Next, we will assume some probability distribution $P(r)$  for the
stripe-stripe distance. For simplicity we will take the ``box''
distribution $P(r)$ centered around $\ell = 1/\delta$ and with finite
width $a = \nu \ell$, where $\nu = {\cal O}(1)$ is a parameter. $P(r) =
C$, for $\ell-a \leq r \leq \ell+a$, and zero otherwise. Here we have
simplified to one length scale for both $a$ and $\ell$. The
normalization constant in 2D is $C = [4\pi \ell a]^{-1}$. In this model
one easily finds  
\beqa 
\langle J(r)
\rangle &=&\int d^2 r P(r) J(r) = {2\pi t_0 C\over{2-\alpha}} a_1
\ell^{2-\alpha} 
\nonumber \ , \\ 
\langle r \rangle &=& {2\pi C\over{3}} a_2
\ell^3 \ , 
\eeqa 
where the constants $a_1,a_2$ are ${\cal O}(1)$. Thus, for $\alpha =
1$, we obtain the experimentally observed relation  
\beqa 
T_c(x) \simeq \langle J(r) \rangle \propto [\langle r \rangle]^{-1} =
\delta(x) \ .
\label{linear} 
\eeqa 
We have examined a variety of distributions $P(r)$ and functional
dependences for $J_{ij}$; Eq.~\ref{ham} with $\alpha =1$ is the only
one reproducing the experimental data (at our mean-field random
Josephson coupling level. Implicit in $J(r)$ is the exponential cutoff
at lengths much larger than the stripe-stripe distance. This cutoff is
necessary to have a well defined thermodynamic limit but is not
important for short length scales). The screening mechanism (magnetic,
elastic fluctuations, etc.) responsible for this form requires detailed
microscopic modeling \cite{antonio}. The present model does not allow
us to determine the magnitude of $v^*$  without making specific
assumptions about parameters such as $t_0$.

In conclusion, we have presented a microscopic model that captures the
essential magnetic and pairing properties of high-temperature cuprate
superconductors. Pairing of holes is a consequence of the existence of
an AF background. (Analogous scenarios in other broken symmetry
backgrounds, e.g. doped charge-density-wave bismuthates, are likely.)
Crucially, however, the glue is provided by magnetic {\it
inhomogeneities} whose precise origin remains to be unraveled, although
it seems fundamental that these perturbations should {\it locally}
break spin-rotational invariance. This pairing mechanism is kinetic
exchange-interaction based and involves a competition between Ising and
$XY$ symmetries. We emphasize that the pair-binding occurs only for
intermediate strengths of $t$ and (local) $J_z$.  We also introduced a
phenomenological model and scenario for the macroscopic
superconductivity based upon coherent Josephson-tunneling of pairs of
holes between these magnetic inhomogeneities in a mesoscopic
liquid-crystal-like \cite{theory} skeleton. We have shown that this
approach is able to recover the magnetic incommensuration $\delta$ and
its experimentally observed relation to $T_c(x)$. Finally, we note that
we have assumed static magnetic inhomogeneities. The case where the
broken spin-symmetry follows the hole is also interesting. Elsewhere,
we will discuss this generalization of coupling the inhomogeneity
selfconsistently to dynamic holes. 

Work at Los Alamos is sponsored by the US DOE under contract
W-7405-ENG-36. 

\begin{figure} 
\epsfxsize=3in 
\caption{Schematic representation of the superconducting ground
state. AF zones between stripe segments are colored in blue. Red spins
near the stripes represent easy-axis Ising-like links. Gray ellipses
characterizes the bound pairs (holes in red). Note the zig-zag alignment
of the holes, and the AF domains $\pi$-shifted at each side of the
stripes.}
\label{fig_binding} 
\end{figure}

\begin{figure} 
\epsfxsize=3in 
\caption{Correlation function for 2 holes in a 16$\times$1 chain with
$d=3$, $P=8$, $J=1$ and $\delta J_{\perp}=-0.9$ (top); and for 4 holes
in 8$\times$2 with  $d=1$, $P=4$ and the other parameters as before
(bottom). Lines in the insets show the perturbed bonds. In 16$\times$1,
as $t$ grows the pair switches from occupying one single stripe, to two
neighboring ones. In the (2 stripes) 8$\times$2 case, the holes remain
at both ends of the anisotropic bond, but as $t$ is increased they form
an effective two-site stripe because of transverse spin fluctuations.
The magnetic energy is lowered by $\pi$-shifting the AF domains on
either side of the stripe.} 
\label{fig_noni} 
\end{figure}

\begin{figure} 
\epsfxsize=3in 
\caption{Spin structure factor for a $t$-$J$$J_z$ ladder (8$\times$2)
with two $\delta J_{\perp}=-0.9$ bonds in the $y$-direction (see
inset). The incommensurability appears only for $t$ larger than a
critical value. When 6 holes are added to the system the double peak
disappears and is replaced by a broad one around ${\bf k}=(0,\pi)$.} 
\label{fig_sq} 
\end{figure}

\begin{figure} 
\epsfxsize=3in 
\caption{Schematic Josephson coupling between an assumed distribution
of stripe segments. For the incommensuration $\delta$ to be observed
along crystallographic (1,0) and (0,1) directions, the stripe-stripe
distances must have average $\langle r \rangle \approx \ell =
1/\delta$. $\langle J \rangle$ will be determined by the probability
distribution $P(r)$ that determines the statistics of inter-stripe
distances. Physically it is clear that $P(r)$ should be centered near
$\ell$, with some width arising from the meandering of stripes (see
text).}
\label{spaghetti} 
\end{figure}

\end{multicols}

\end{document}